\documentclass[pra,aps,twocolumn,superscriptaddress,longbibliography]{revtex4-2}

\usepackage{amsmath, graphicx, color, times, float, amssymb,xcolor}
\usepackage[colorlinks=true,citecolor=blue,linkcolor=blue,urlcolor=blue]{hyperref}
\usepackage{url}
\usepackage{comment}

\usepackage{soul}

\newif\ifshowcomments
\showcommentstrue   

\begin{document}

\title{Dynamical Signatures and Kibble–Zurek Scaling of Localization in Tilted Bose-Einstein Condensates}

\author{Argha Debnath}
\affiliation{Harish-Chandra Research Institute, Chhatnag Road, Jhunsi, Allahabad 211 019, India\\
Homi Bhabha National Institute, Training School Complex, Anushakti Nagar, Mumbai 400 094, India}

\author{Radheshyam Saha}
\affiliation{Harish-Chandra Research Institute, Chhatnag Road, Jhunsi, Allahabad 211 019, India\\
Homi Bhabha National Institute, Training School Complex, Anushakti Nagar, Mumbai 400 094, India}

\author{Ravindra W. Chhajlany}
\affiliation{Faculty of Physics, Adam Mickiewicz University, 61614 Poznan, Poland}

\author{Mariusz Gajda}
\affiliation{Institute of Physics, Polish Academy of Sciences,
Aleja Lotnikow 32/46, PL-02668 Warsaw, Poland}

\author{Debraj Rakshit}
\affiliation{Harish-Chandra Research Institute, Chhatnag Road, Jhunsi, Allahabad 211 019, India\\
Homi Bhabha National Institute, Training School Complex, Anushakti Nagar, Mumbai 400 094, India}

\begin{abstract}
We study nonequilibrium signatures of tilt-induced localization in a one-dimensional Bose-Einstein condensate loaded in a shallow optical lattice. The tilt strength acts as a control parameter for the localization-delocalization crossover. We also consider the effects of repulsive interactions, which tend to delocalize the condensate. We first characterize localized and delocalized regimes through sudden quenches of the interaction strength and the external tilt. The resulting dynamics is analyzed using the survival probability and its power spectral density. Localized condensates exhibit strong memory retention, pronounced revivals, regular dynamics and a narrow spectral response, whereas delocalized condensates show suppressed recurrences, irregular dynamics and a broader distribution of spectral weight over many frequencies. We then investigate finite-rate ramps of the tilt strength across the localization threshold. Using the localization length and the Bogoliubov excitation gap, we extract the relevant critical exponents and perform Kibble-Zurek scaling analysis in the driven dynamics. Our results establish quench response and finite-rate scaling as complementary dynamical probes of localization in interacting Bose gases, with direct relevance to cold-atom experiments in tilted optical lattices.
\end{abstract}

\maketitle

\section{\label{sec:intro}Introduction}

Localization -- a transport inhibiting phenomena, introduced by Anderson \cite{anderson_loca} and pursued by many \cite{billy2008direct,roati2008anderson,cheng2010matter,xi2015localization,white2020observation,zhang2022anderson}  -- is not solely restricted to  disorder. Complementary routes to achieve localization in the absence of disorder can be sought through  quasi-periodic \cite{aubry1980analyticity,modugno2009exponential,adhikari2009localization,muruganandam2010localization,cheng2010symmetry,cheng2010spatially,cheng2010spatially,cheng2011localization,cheng2011matter,iyer2013many,michal2014delocalization,cheng2014localization,li2016localization,modak2021many} or tilted potential \cite{fukuyama1973tightly, kolovsky2008interplay, kolovsky2013wannier,van2019bloch, schulz2019stark,bhakuni2020drive, taylor2020experimental,morong2021observation,yao2021many,wei2022static}. On the other hand many-body localization (MBL) deals with the fate of localization when multiple interacting particles are subjected to these potential landscapes \cite{Abanin,altman2018many,ALET2018498,abanin2019colloquium,Moudgalya_2022}. Interaction, in general, acts as a delocalizing agent and its influence on localization has been a subject of intense research in recent times \cite{pikovsky2008destruction,kopidakis2008absence,deissler2010delocalization,lucioni2011observation,sarkar2023quench}. 

The localization transition can be tracked by probing
localization length, inverse participation ratio, energy gap, fidelity susceptibility and their associated scaling exponents are 
standard tools 
for investigating this phenomena 
\cite{vojta2003quantum,Sachdev_2011,bu2022quantum,sahoo2025stark,debnath2025localization,argha_tilt}. 
Apart from these static properties, out of equilibrium dynamics across the localization transition 
serves to distinguish between 
localized and delocalized states. Localization behaviour means that an initially prepared localized state remains spatially confined 
while delocalization is assocated with spreading across the system -- this manifests in differing behaviour of  transport, coherence, and relaxation properties.

Ultra-cold atomic gases loaded into optical lattices (OL) \cite{cristiani2002experimental,lewenstein2007ultracold} are a highly controllable playground for proving several intricate aspects of localization. 
Interestingly, the dynamical evolution of Bose–Einstein condensate (BECs) upon
interaction or potential quenches may bear signatures of periodic, aperiodic or chaotic motion \cite{muruganandam2002chaotic,sanchez2007anderson,adhikari2009localization,bvrezinova2011wave,lucioni2011observation,cheng2011localization,van2019bloch,doggen2014quench,tosyali2018regular,du2022quench,sarkar2023quench}. 
The opposite limit of slow modulation of system paramaters on quantum dynamics has also garnered significant interest recently. 
In particular, the Kibble–Zurek (KZ) mechanism provides a general prescription that describes universal behaviour of  defect production  associated with driving a system across a continuous phase transition  by slow (finite rate) control parameter ramps \cite{kibble1976topology,kibble2007phase,zurek1985cosmological,zurek1996cosmological,zurek2005dynamics,polkovnikov2005universal,dziarmaga2010dynamics,polkovnikov2011colloquium, sinha2019kibble,bu2022quantum,bu2023kibble,zhai2022nonequilibrium,liang2024quantum,wang2025driven,debnath2025localization}. The KZ mechanism has been investigated in a wide range quantum systems, particularly in quantum gases and 
theoretically \cite{schutzhold2006sweeping,damski2007dynamics,uhlmann2007vortex,del2011inhomogeneous,sen2008defect,su2013kibble,beugnon2017exploring,liu2019quench,comaron2019quench,jiang2019universality,liu2020kibble,thudiyangal2024universal,wheeler2025dynamics,kirkby2025kibble}
and experimentally \cite{weiler2008spontaneous,lamporesi2013spontaneous,corman2014quench,navon2015critical,chomaz2015emergence,donadello2016creation,clark2016universal,anquez2016quantum,feng2018coherent,liu2018dynamical,goo2021defect,goo2022universal,rabga2023variations,chen2019dynamical,yi2020exploring}.  In contrast, sudden quenches probe the intrinsic stability and relaxation properties of the resulting phases, revealing information regarding recurrence, memory retention, and spectral complexity.

A recent investigation on BEC loaded to a tilted OL has shown a great promise of witnessing localization–delocalization transitions governed by the competition between trapping potential, kinetic energy, and nonlinear interactions \cite{argha_tilt}. The tilt strength acts as a localization control parameter, while repulsive interactions compete against localization and promote delocalization. While static properties of these transitions have been explored, this work perform studies on its nonequilibrium extensions. In this work, we investigate two separate aspects of non-equilibrium dynamics in a one-dimensional BEC loaded to a lilted OL.  

To establish a complementary dynamical characterization of tilt-induced localization transitions in interacting Bose gases -- First, we analyze post-quench dynamics through survival probability and power spectral density following sudden changes in interaction strength or external tilt. We show that localized condensates exhibit persistent revivals and periodic dynamics, whereas delocalized condensates display increasingly irregular temporal evolution. 

Second, we characterize the localization transition through finite-size scaling of the localization length and Bogoliubov excitation gap, allowing extraction of the critical exponents associated with the transition. Using these exponents, we study the dynamics due to slow ramping across the critical point and demonstrate universal
Kibble–Zurek scaling (KZS). Moreover, it is shown that KZ mechanism provides an independent dynamical verification of the critical exponents extracted from the static analysis.

The motivation of the present work is to understand whether localization induced by an external tilt leaves identifiable signatures in nonequilibrium dynamics. Universal scaling during slow ramps and memory retention following sudden quenches probe distinct physical mechanisms. While KZ  dynamics characterizes the emergence of excitations due to critical slowing down, quench dynamics measures the ability of localized states to preserve coherence and resist spreading. Investigating both within a common framework provides a more complete picture of localization transitions beyond static properties. Together, the KZ  analysis and quench dynamics provide complementary perspectives on localization.

The structure of this paper is organized as follows. In
Sec.~\ref{sec:model}, we present the governing equations and parameters responsible for localization-delocalization transition. In Sec.~\ref{sec:quench_dynamics} we discuss quench dynamics and calculate survival probability and power spectral density to distinguish dynamical behavior of localize and delocalize states. In Sec.~\ref{sec:kzs} we investigate KZ dynamics by ramping tilt strength from localized to delocalized state. 
Finally, we present our conclusions in Sec.~\ref{sec:conc}.

\section{\label{sec:model} Model For Tilt-Induced Localization}
We consider a dilute and weakly interacting BEC trapped in a one-dimensional potential  with
strong transverse confinement, such that the transverse degrees of freedom remain frozen, and describe 
this quasi-one-dimensional BEC within the mean-field approximation by the Gross–Pitaevskii equation (GPE). After appropriate dimensional reduction under strong transverse confinement, the condensate dynamics is governed by \cite{argha_tilt}
\begin{eqnarray}
i \partial_t \psi=(-\partial^2_x/2+V_{ext}(x)+g|\psi|^2)\psi,\label{eq:one}
\end{eqnarray}
where $\psi$ is a complex scalar field representing the interacting BEC order parameter  with $\int dx |\psi(x,t)|^2 = 1$. $g$ represents the effective nonlinear interaction strength arising from repulsive atom–atom interactions. Experimentally, $g$ can be tuned  over a broad range by either controlling the atom number or by tuning the scattering length via magnetic field near a Feshbach resonance~\cite{inouye1998observation}.
Eq.~\ref{eq:one} contains dimensionless variables introduced as : $x \to x k$,  $t \to 2E_k t/\hbar$, $g\to N \frac{2 \hbar^2 a_s}{m E_k a_{\perp}^2}\left(1-C a_s/a_{\perp}\right)^{-1}$ and $\psi \to \psi/\sqrt{k} $. The unit of distance $1/k$ is related to the period of the external
lattice potential $V_{ext}(x)$ (see Eq.~\ref{eq:two}). Here $E_k = \hbar^2 k^2/2m$ is the recoil energy, where $m$ is the atomic mass, $a_s$ is the 3D scattering length, $a_{\perp}$ is the transverse confinement length, and $C \approx 1.4603$ \cite{olshanii1998atomic}. 

The potential is represented by $E \sin^2(kx)$ and superimposed on the V-shaped trap $f|x|$, where $f$ is a constant force. The BEC experiences the effect of a combined potential
\begin{equation}
V_{\text{ext}}(x) = V \sin^2(kx) + V_0 |x|.
\label{eq:two}
\end{equation}
Here  $V=E/(2E_k)$, $V_0=f/(2kE_k)$. This system can be viewed as two Stark-type tilted lattices with opposite slopes joined at a cusp at 
$x=0$. 

The protocols considered here are also directly motivated by cold-atom experiments. Tilted OL potential is experimentally realizable. Particularly, a V-shaped or symmetric linear Stark potential can be realized in experiments using magnetic quadrupole trapping fields along with the interference of counter-propagating two linearly polarized laser beams with associated wavevector $k$.  The interaction strength can be tuned using the atom number or scattering length. 

 We impose the kinetic energy constraint $\kappa^2 = 2 (V- \mu) < 1$ \cite{efremidis2003lattice}, where $\mu$ is the chemical potential, to ensure that the system is away from the tight-binding limit and it operates within a continuum-like regime amenable to the treatments via GPE. In the absence of an applied tilt (i.e., with  $V_0=0$) and in a weakly modulated OL, the GPE admits extended stationary states analogous to Bloch waves, which span the entire optical lattice and reflect its translational symmetry. We consider 
 a shallow optical lattice with depth $V=0.5$ and the potential ratio $V_0/V$ is denoted by $\alpha$. As shown in Ref.~\cite{argha_tilt} this setup enables an isolated investigation of how a linear potential (the tilt) affects localization–delocalization behavior in the presence of a shallow OL. The GP description  provides a successful and an effective means for studying the static physics localization due to the applied tilt. The finite-size systems for a given interaction are characterized by delocalized phase arising at weak tilt strengths $\alpha=0$ and by a localized phase that emerges beyond certain finite value, say $\alpha_c$, marking the crossover between two distinct phases in these finite-size systems.  At the same time, for a localized state, interaction $g$ acts as a delocalizing agent. Repulsive interaction influences these localized states and significantly suppresses the emergence of localization at values of $\alpha$ where localization would occur in the absence of interactions. In the following, we analyze the characteristics of these states using their dynamical evolution.

\section{\label{sec:quench_dynamics}Characterizing the Dynamics of Localized and Delocalized States}
To characterize the nonequilibrium evolution of the condensate, we analyze the survival probability, or fidelity, quantified as the overlap between the initial state $\psi(x,0)$ and the time evolved state $\psi(x,t)$. It is defined as
\begin{equation}
SP(t)=\left|\int_{-\infty}^{\infty}\psi^*(x,0)\psi(x,t)\,dx\right|^2.\label{eq:three}
\end{equation}
The survival probability quantifies the degree of memory retained by the system during time evolution. Persistent revivals of $SP(t)$ after fixed time intervals indicate regular dynamics, whereas irregular temporal variations suggest the involvement of multiple competing frequencies. To further resolve the underlying dynamical behavior, we compute the power spectral density (PSD) of $SP(t)$ \cite{dalui2020induction,sarkar2023quench},
\begin{equation}
PSD=\frac{1}{2\pi \mathcal{N}}
|\hat{SP}(\omega_n)|^2,
\label{eq:four}
\end{equation}
where $\hat{SP}(\omega_n)$ denotes the discrete Fourier transform of $SP(t)$ and is defined as

\begin{equation}
\hat{SP}(\omega_n)=
\sum_{m=0}^{\mathcal{N}-1}
SP(t_m)\,
e^{-i\omega_nt_m},\label{eq:fft}
\end{equation}
where $SP(t_m)$ is sampled at times $t_m=m\,dt$ $(m=0,1,...,\mathcal{N}-1)$,
with $dt$ being the evolution time step. $\omega_n=\frac{2\pi n}{\mathcal{N}\,dt}$, where $n$ is the index labeling the discrete frequencies. The PSD provides information on the dominant frequencies governing the evolution and allows one to distinguish periodic, quasiperiodic, and chaotic dynamics.

\subsection{Quench Dynamics Of The Localized And Delocalized States}
We perform dynamical studies in order to understand the effect of $\alpha$ ($g$) by keeping $g$ ($\alpha$) fixed. Throughout the dynamical analysis we consider the BEC to be confined in a finite OL of size $L=60: x\in [-30, 30]$, and it is subjected to a closed boundary condition with $\psi(\pm30)=0$, unless mentioned otherwise.

\begin{figure}
\includegraphics[scale=0.30]{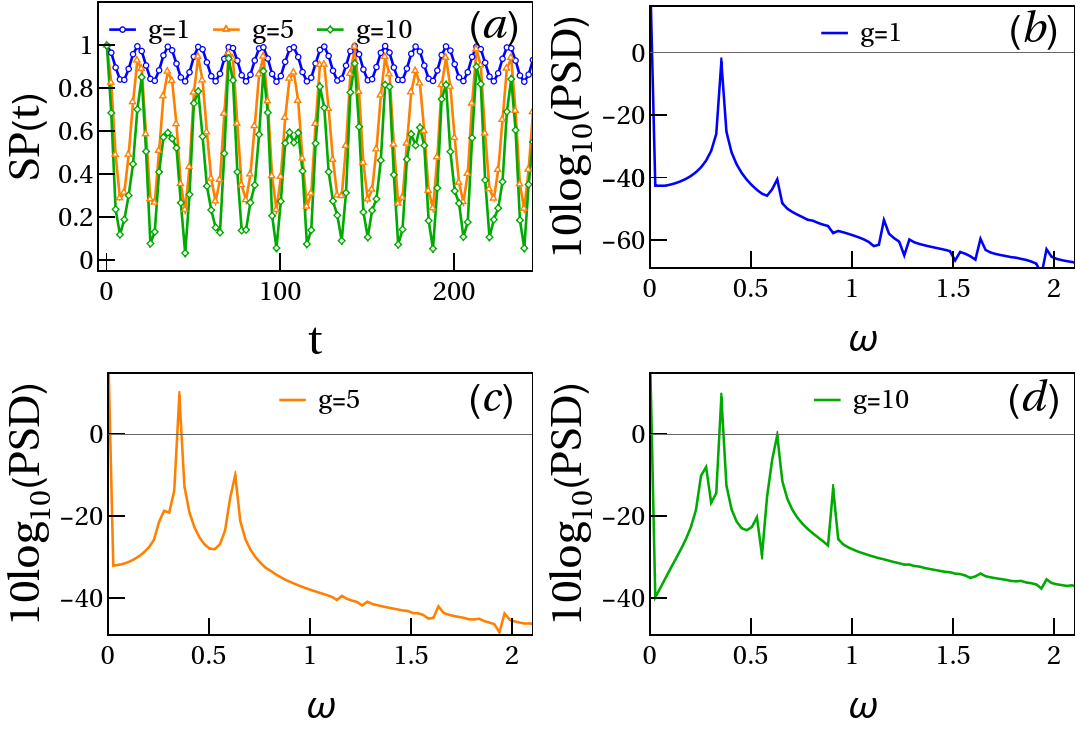}
\caption{(a) Temporal evolution of the survival probability $SP(t)$ at $\alpha=0.2$ after quenching the interaction strength from an initial value $g$ to zero. The initial states are prepared as ground states for $g=1$ (blue circles), $g=5$ (orange up-triangles), and $g=10$ (green diamonds). The localized phase ($g=1$) displays regular periodic oscillations with strong revivals, while increasing interaction-induced delocalization ($g=5$ and $10$) gives rise to quasiperiodic and increasingly irregular dynamics. The corresponding PSD of $SP(t)$ is shown in (b)–(d) for $g=1$, $5$, and $10$, respectively.}\label{fig:interaction_quench}
\end{figure}

\subsubsection{Interaction Quench Dynamics}
To study the detailed dynamics of the localized and delocalized condensates, we consider the ground states obtained for different values of nonlinearity $g$ and perform the time evolution by applying an instantaneous quench of the nonlinear interaction to zero. The resulting dynamics is obtained by real-time evolution of the GPE. To probe the spatiotemporal evolution after quenching of the condensate prepared at different, we investigate the dynamics of the localized matter wave density by computing the survival probability $SP(t)$ defined in Eq.~\ref{eq:three}, and analyze its temporal evolution. For this, we consider  a number of the ground states $\psi(x,0)$, obtained for
increasing values of $g$,  as the initial condensate wave functions and
monitor their subsequent time evolution after quenching interactions to
zero, $g=0$. Fig.~\ref{fig:interaction_quench}(a) shows the survival probability $SP(t)$ for condensates initially prepared at $g=1$ (blue circle), $5$ (orange up-triangle ) and $10$ (green diamond), while keeping $\alpha=0.2$ fixed. For $g=1$ (blue circle), corresponding to a localized condensate, $SP(t)$ exhibits near periodic oscillations dominated by a single frequency, implying strong memory retention of the initial state. Increasing the interaction to $g=5$ (orange up-triangle) introduces amplitude modulation in $SP(t)$, suggesting the coexistence of multiple frequencies and quasi-periodic behavior. For stronger interaction $g=10$ (green diamond), the amplitudes of oscillations
display irregular temporal structures, indicating delocalized nature of the initial condensate. This signifies a gradual transition from periodic to increasingly complex dynamics as interactions promote delocalization.

The PSD (see Eq.~(\ref{eq:four})) analysis provides a deeper understanding of the observed behavior of $SP(t)$. $g=1$ case in Fig.~\ref{fig:interaction_quench}(b) is characterized by a dominant peak at $\omega \sim 0.35$, characteristic of regular oscillation. Fig.~\ref{fig:interaction_quench}(c) displays the case of $g=5$, which is marked by an additional peak near $\omega \sim 0.63$, reflecting quasiperiodic dynamics involving more than one frequency scale. For $g=10$, several dominant peaks appear at $\omega_1 \sim 0.276$, $\omega_2 \sim 0.35$, $\omega_3 \sim 0.53$, $\omega_4 \sim 0.63$ and $\omega_5=0.9$ (see Fig.~\ref{fig:interaction_quench}(d)). These frequencies can be approximately interpreted as nonlinear combinations of two principal frequencies, for example $\omega_1+\omega_2\approx\omega_4$
and $2\omega_1+\omega_2\approx\omega_5$. The enhanced PSD peaks of mixed frequencies signal progressive complex dynamics due to the interaction-induced delocalization. This brings us to conclude that interaction quenches reveal a progression from periodic oscillations in localized states to quasiperiodic and irregular dynamics as repulsive interaction delocalizes the condensate.

\subsubsection{Tilt-Induced Localization Quench}
\begin{figure}
\includegraphics[scale=0.30]{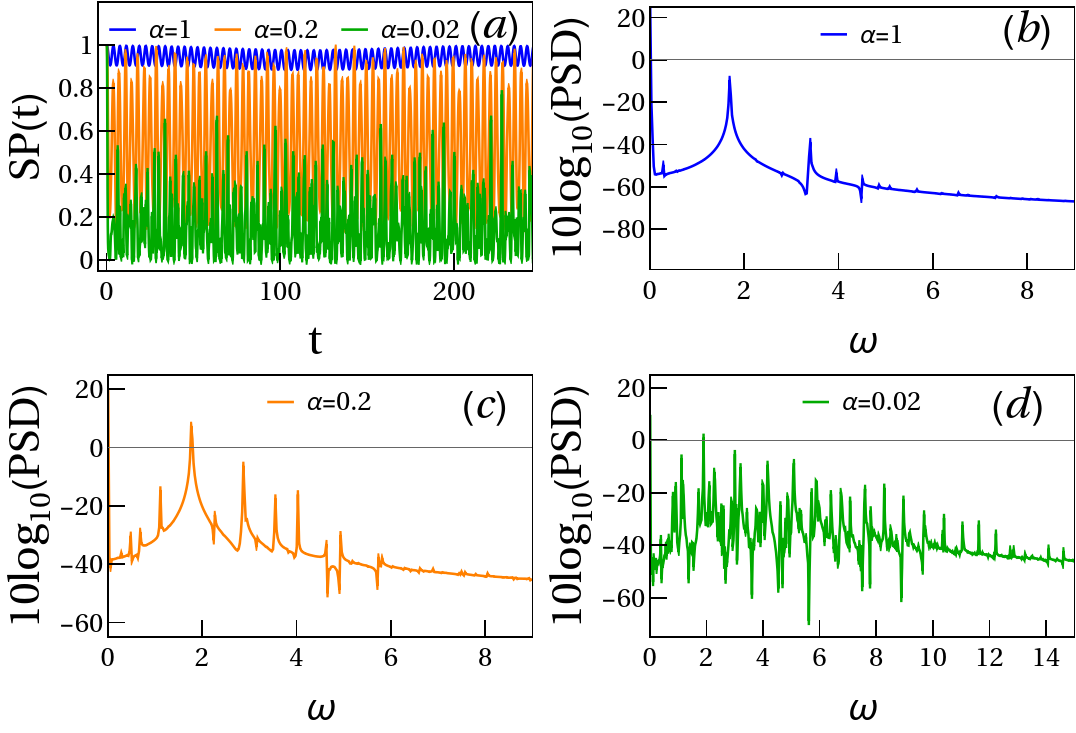}
\caption{(a) Temporal variation of the survival probability $SP(t)$
at $g=1$ after quenching $\alpha$ from an initial value to $2$. The initial states are prepared as ground states for $\alpha=1$ (blue), $\alpha=0.2$ (orange), $\alpha=0.02$ (green). In the localized state ($\alpha=1$), $SP(t)$ exhibits regular periodic oscillations with strong revivals, which becomes irregular in the delocalized state corresponds to smaller $\alpha$ ($\alpha=0.2$ and $0.02$). The corresponding PSD of $SP(t)$ is shown in (b)–(d) for $\alpha=1$, $\alpha=0.2$, $\alpha=0.02$, respectively.}\label{fig:tilt_quench}
\end{figure}

Next, we perform an analogous investigation for initial states corresponding to different values of the tilt strengths by fixing the interaction strength. Ground states are prepared at different $\alpha$, and then evolved after abruptly changing it to a fixed final $\alpha$, say $\alpha=\alpha_f$. Fig.~\ref{fig:tilt_quench}(a) shows $SP(t)$ for initial states prepared at $\alpha=1$ (blue), $0.2$ (orange), and $0.02$ (green) with the final tilt of the post-quenched Hamiltonian $\alpha_f=2.0$. For $\alpha=1$, corresponding to a localized condensate, $SP(t)$ exhibits oscillation with a single dominant frequency, indicating strong memory retention of the initial state. Reducing the tilt to $\alpha=0.2$ leads to strongly modulated oscillations, while for $\alpha=0.02$ the dynamics become highly irregular with suppressed revivals.

The PSD further highlights these features. For $\alpha=1$, Fig.~\ref{fig:tilt_quench}(b) exhibits a dominant frequency
$\omega_1=1.68$ and its harmonic 
$\omega_2=3.39\approx2\omega_1$, consistent with periodic evolution. At $\alpha=0.2$, numerous frequencies appear ($0.48, 0.68, 1.11, 1.76, 2.26, 2.86, 3.54, 4.02, 4.62, 4.92$) (see Fig.~\ref{fig:tilt_quench}(c)), producing a broadened spectrum that cannot be expressed simply through harmonic combinations of a few modes. For the weakly tilted case $\alpha=0.02$, Fig.~\ref{fig:tilt_quench}(d) shows a dense distribution of spectral peaks extending over a broad frequency range, reflecting irregular dynamics associated with delocalized condensates. These results demonstrate that stronger tilt-induced localization stabilizes regular oscillatory dynamics, whereas reducing the tilt promotes increasingly complex evolution.

The quench dynamics discussed above establishes that localization can not only be probed via static means \cite{argha_tilt}, but also in the way the condensate responds to perturbations. Survival probability and its power spectrum can effectively captpture the distinct dynamical fingerprints of localized or delocalized states: strong revivals and narrow spectral response characterize localized condensates, whereas delocalized condensates display irregular temporal behavior and spectral broadening. This naturally raises a complementary question: how does the system dynamically pass between these two regimes when the localization control parameter is varied continuously? Near the localization threshold, the localization length grows and the relevant Bogoliubov excitation gap is suppressed, indicating a slowing down of the intrinsic response. We now turn from sudden quench to slow parameter ramps, where the central question is how adiabatic following breaks down as the localization threshold is approached. This leads naturally to the Kibble-Zurek (KZ) analysis presented below.

\section{\label{sec:kzs}Critical Scaling and Kibble-Zurek mechanism}
Although the KZ mechanism was formulated originally for continuous symmetry-breaking phase transitions, its scaling structure relies more generally on the presence of a diverging length scale and a vanishing characteristic energy scale. Localization transitions provide a natural setting for such an extension. Here, the localization length replaces the correlation length as the relevant diverging scale, while the Bogoliubov excitation gap determines the intrinsic time scale for adiabatic response. Consequently, when the tilt strength is ramped across the localization threshold, the system is expected to depart from adiabatic evolution in the critical region and exhibit universal finite-rate scaling. This motivates the KZ analysis of the driven dynamics presented below. We show that the universal behavior can be described by a small set of critical exponents, $\nu$ and $z$, describing the power-law divergence of the localization length, which here we characterize by root-mean square (RMS) width,  and the vanishing of the Bogoliubov excitation gap, respectively.

\subsection{\label{sec:static scaling}Static Scaling of Localization Length and Excitation Gap}
In order to extract critical exponents, $\nu$ and $z$, we need to understand the dependence of localization length and energy gap on system size. In the following, we study the ground state behavior of these quantities as  $\alpha$ is varied for different system sizes.
\subsubsection{\label{subsec:loca}Localization Length}
In order to formally understand the nature of the  localization transition and associated critical properties in the thermodynamic limit, we compute RMS width, $\xi$, which is defined as,
\begin{eqnarray}
\xi^2=\int_{-\infty}^{\infty} (x-\langle x\rangle)^2|\psi(x,t)|^2 dx,\label{eq:five}
\end{eqnarray}
where $\langle x \rangle=\int_{-\infty}^{\infty} x|\psi(x,t)|^2 dx$. The static analysis of $\xi$ for a finite interaction strength $(g=1)$ on the control parameter $\alpha$ for different system sizes $L$  in Ref.~\cite{argha_tilt}. The RMS width follows the relation $\xi\propto |\alpha-\alpha_c|^{-\nu}$, where the scaling exponent $\nu$ determines the rate of divergence of the RMS width in the thermodynamic limit near criticality and $\alpha_c$ is the critical strength of the tilt corresponding to the localization in the thermodynamic limit. The scaling exponent, $\nu$, and $\alpha_c$ can be obtained via data collapse adopting the following scaling ansatz,
\begin{eqnarray}
\xi=&&L f_1\left((\alpha-\alpha_c)L^{1/\nu}\right),\label{eq:six}
\end{eqnarray}
where $f_1[.]$ is an arbitrary function. 
The numerically obtained scaling exponents from the static analysis are reported as $(\nu, \alpha_c) \sim \left(0.42, 0.00027\right)$ \cite{argha_tilt}. 

\subsubsection{\label{subsec:energygap}Small-Amplitude Oscillations}
As in usual quantum criticality, the energy gap between the first excited state and the ground state can also be used to characterize the criticality. 
While $\nu$ has already been investigated in Ref.~\cite{argha_tilt}, below we perform an explicit static analysis for exacting the scaling exponent associated with the energy gap.

The collective behavior exhibited by the interacting Bose gases can be interpreted in terms
of the elementary excitations of the system governed by GPE. The small-amplitude oscillations can also be interpreted in terms of the elementary excitations of
the system and they admit a natural quantum description. For describing the elementary excitation of a BEC, the Bogoliubov theory starts with the ground state $\psi_{gs}(x)$ of GPE and assumes the evolution of GPE around the stationary state which takes the form of 
\begin{eqnarray}
\psi(x,t)=e^{-i\mu_{gs} t}\left[\psi_{gs}(x)+\mathcal{\vartheta}(x,t)\right],\label{eq:seven}
\end{eqnarray}
where $\mu_{gs}$ is the ground state chemical potential and $\mathcal{\vartheta}$ is a small variation for which we look for solutions of the form
\begin{eqnarray}
\mathcal{\vartheta}(x)=\sum_i\left[u_i(x) e^{-i\Omega_i t}-v_i(x) e^{i\Omega_i t}\right],\label{eq:eight}
\end{eqnarray}
where $\Omega_i$ is the frequency of the oscillation. The functions $u_i(x)$ and $v_i(x)$ are determined by solving the GPE in the linear limit. By collecting all the terms evolving in time like $e^{\pm i\Omega_i t}$, we obtain the following pair of coupled set of differential equations (the  so-called Bogoliubov equations): 
\begin{eqnarray}
\Omega_i u_i(x) &=& [-\partial^2_x/2 + V \sin^2(x) + V_0 |x| + 2g|\psi_{gs}(x)|^2\nonumber\\ && \ -\mu_{gs}] u_i(x)-g(\psi_{gs}(x))^2 v_i(x),\nonumber \\
-\Omega_i v_i(x) &=& [-\partial^2_x/2 + V \sin^2(x) + V_0 |x| + 2g|\psi_{gs}(x)|^2 \nonumber\\ && -\mu_{gs}] v_i(x)- g(\psi_{gs}^*(x))^2 u_i(x),
\end{eqnarray}

This is a generalized eigenvalue problem, which can be written as
\begin{equation}
 \begin{pmatrix}
    H+g|\psi_{gs}(x)|^2 & -g(\psi_{gs}(x))^2\\
    g(\psi_{gs}^{*}(x))^2 &  -H-g|\psi_{gs}(x)|^2\\
\end{pmatrix}
 \begin{pmatrix}
 u(x)\\
 v(x)
 \end{pmatrix}
= \Omega\begin{pmatrix}
 u(x)\\
 v(x)
 \end{pmatrix}
\end{equation}
under the normalized condition $\int|u(x)|^2-|v(x)|^2 dx=1$ and $\Omega$ is the eigen-frequency, $(u,v)$ is the corresponding eigen-vector.
$H=-\partial^2_x/2 + V \sin^2(x) + V_0 |x| + g|\psi_{gs}(x)|^2 - \mu_{gs}$. Here, the relevant ones are the positive eigenvalues $\Omega_i\geq 0$, or more specifically, the lowest eigenvalue $\Omega_1>\Omega_0$, where $\Omega_0\approx0$. It is important to note that real $\Omega_i$'s correspond to stable solution under small perturbation, and on the other hand, imaginary $\Omega_i$'s lead to instability.
\begin{figure}
\includegraphics[scale=0.35]{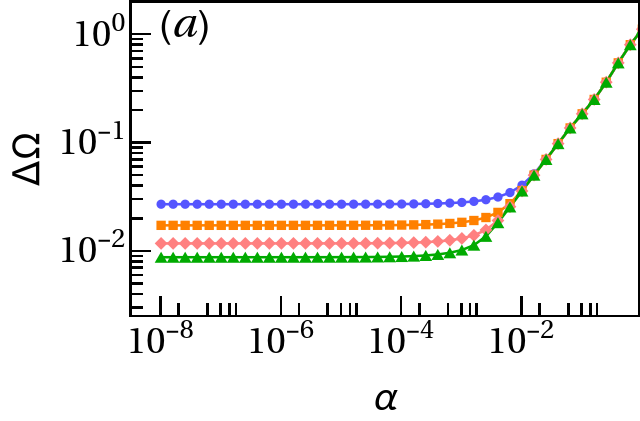}
\includegraphics[scale=0.35]{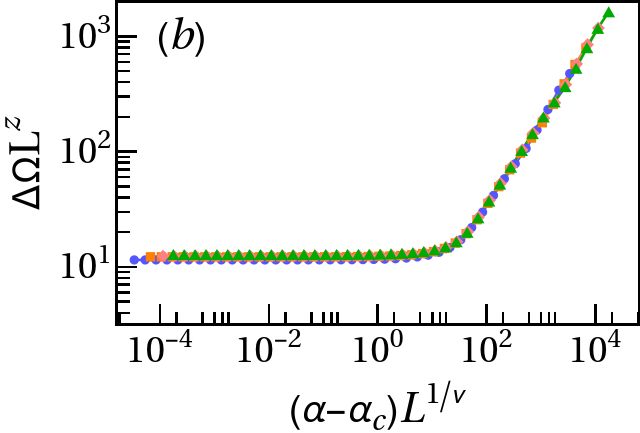}
\caption{Variation of (a) energy gap($\Delta \Omega$) with  $\alpha$ for different system sizes, $L=30$ (blue circle), 40 (orange up-triangle), 50 (pink diamond), 60 (green square) for groundstate. Collapse plot with $(\nu,z)\sim(0.42,1.78)$ are shown for (b) $\Delta \Omega$ with their re-scaled axis.}\label{fig:bdg}
\end{figure}

Fig.~\ref{fig:bdg}(a) shows the energy gap $\Delta \Omega$ with  $\alpha$ for different system sizes. We define $\Delta \Omega=\Omega_1-\Omega_0$. A very narrow energy gap in the delocalized phase becomes wider in the localized phase. While in the delocalized regime, the system has a super-extensive dependence on the system size; it changes drastically in the localized regime, and $\Delta \Omega$ becomes nearly independent in the localized phase. The trend remains similar to what has been reported in context of the localization length \cite{argha_tilt}. To extract scaling function, we consider the following ansatz,
\begin{equation}
\Delta \Omega=L^{-z}f_2\left((\alpha-\alpha_c) L^{1/\nu}\right),\label{eq:eng_an} 
\end{equation}
where $f_2[.]$ is an arbitrary function.
Using the above ansatz, we perform data collapse of the quantity $\Delta \Omega$ and plot it in Fig.~\ref{fig:bdg}(b), with their corresponding scaled axis. The obtained scaling exponents are $\{\nu, z\}=\{0.42,1.78\}$. Note that the value of $\nu$ obtained from the energy gap is the same as the one determined by studying RMS width.

\subsection{Kibble-Zurek Mechanism}

 We now turn to the investigation of Kibble-Zurek Scaling (KZS) in the driven dynamics of our model, which is intrinsically connected to the quantum criticality of the phase transitions. The KZ mechanism establishes a unified framework that explains how systems produce excitations when they experience continuous phase transitions at a controlled speed. The system maintains adiabaticity when the control parameter remains distant from the criticality. The system generates unavoidable excitations because the essential energy gap approaches zero as it approaches the critical point. The KZ framework describes how adiabaticity fails during a process which starts from an easily accessible ground state before reaching a complex final state. The system is initially prepared in the localized phase and is subsequently driven across the critical point by linearly ramping of the control parameter. We linearly vary $\alpha$ in time $t$ with speed $R$. The time evolution of $\alpha$ is given by
\begin{eqnarray}
\alpha(t)=\alpha_i+(\alpha_f-\alpha_i)Rt,\label{eq:kz}
\end{eqnarray}

\begin{figure}
\includegraphics[scale=0.35]{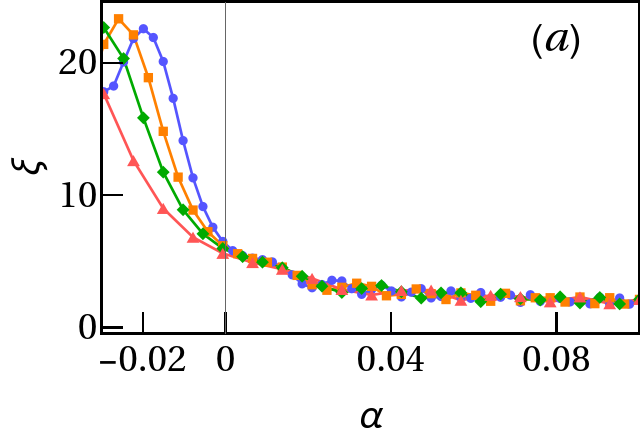}
\includegraphics[scale=0.35]{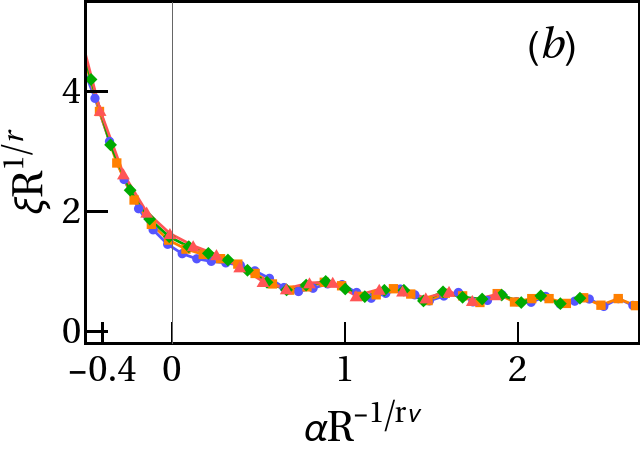}
\caption{Driven dynamics with the initial
state being the ground state. The curves of $\xi$ versus $\alpha$ for $R=0.002$ (blue circle), $R=0.003$ (orange square), $R=0.004$ (green diamond), $R=0.006$ (pink down-triangle) (a) before and (b) after rescaled with $R$.}\label{fig:kz}
\end{figure}

where $\alpha_i>0$ represents the initial distance from the critical point at $t = 0$. Within the KZS framework, adiabaticity can be maintained when the condition  $|\alpha|>R^{1/r\nu}$ is satisfied, with the scaling exponent $r=z+1/\nu$. In this regime, the system can adapt to the gradual variation of the Hamiltonian and thus follow the instantaneous ground state. By contrast, when $|\alpha|<R^{1/r\nu}$, the intrinsic response of the system becomes slower than the external driving rate, signaling its entry into the impulse regime, where the evolution effectively freezes and excitations are generated.

To see how $\xi$ behaves around criticality while driven dynamically, we introduce following ansatz,
\begin{equation}
\xi=R^{-1/r}f_3\left(\alpha R^{-1/r\nu}\right), \label{eq:kz_an1}
\end{equation}
where $f_3[.]$ is arbitrary function. We present the results with moderate system size, $L=60$. We set the initial state as the ground state of the system with $\alpha_i=0.2$ and $\alpha_f=-0.04$. Fig.~\ref{fig:kz}(a) shows the evolution of
the localization length $\xi$ for different $R$. Initially, a strongly localized state with a small $\xi$ remains almost insensitive to the ramp rate $R$, allowing the dynamics to closely follow the instantaneous ground state in an effectively adiabatic manner. However, as $\alpha$ approaches to the critical point, the trajectories corresponding to different ramp rates $R$ begin to separate, signaling the onset of the impulse regime where the evolution can no longer remain adiabatic. We rescale $\xi$ and $\alpha$ as $\xi R^{1/r}$ and $\alpha R^{-1/\nu r}$, respectively, according to the ansatz in Eq.~\ref{eq:kz_an1}.  We find that the rescaled curves collapse onto each other near the critical point, as shown in Fig.~\ref{fig:kz}(b), with the obtained scaling exponents from the static case. 

\begin{figure}
\includegraphics[scale=0.35]{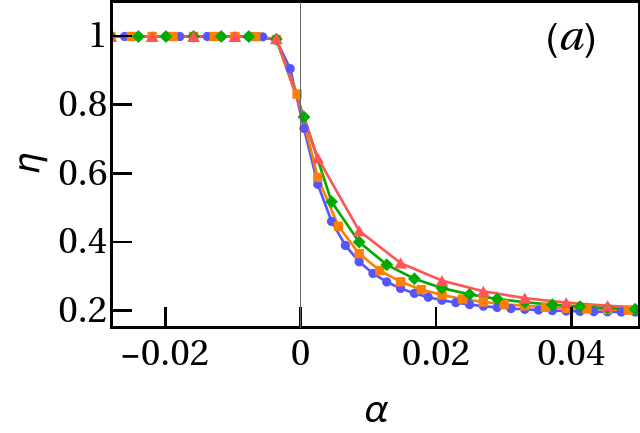}
\includegraphics[scale=0.35]{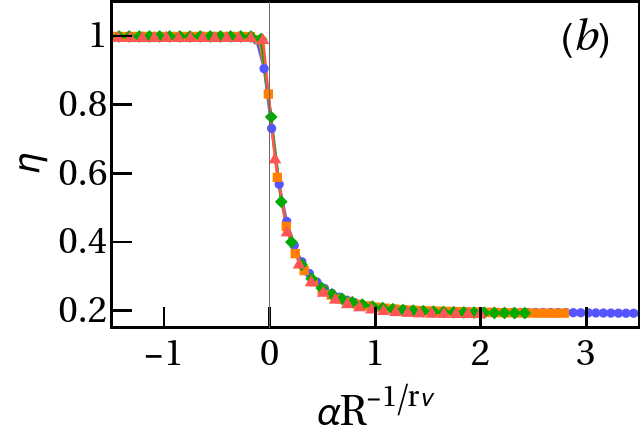}
\caption{Driven dynamics with the initial
state being the ground state. The curves of $\eta$ versus $\alpha$ for $R=0.002$ (blue circle), $R=0.003$ (orange square), $R=0.004$ (green diamond), $R=0.006$ (pink down-triangle) (a) before and (b) after rescaled with $R$.}\label{fig:kz_defect}
\end{figure}

KZ introduces excitations thus dynamical deviation from the instantaneous ground state is an important quantity  in characterizing the dynamical behavior of localization phase transitions. In order to investigate this we evaluate following quantity,
\begin{equation}
\eta=1-|\langle\psi_{gs}(\alpha(t))|\psi_{KZ}(\alpha(t))\rangle|^2. \label{eq:kz_an2}
\end{equation}
Away from critical point $\eta$ remains nearly flat, implying that the evolved state resembles the corresponding ground state while maintaing a large overlap (see Fig.~\ref{fig:kz_defect}(a)). However, near criticality  excitations come into play and adiabaticity breaks down. 
As expected, any physical state of a system, initially in equilibrium, will necessarily become excited when approaching a continuous phase transition.  To see how $\eta$ behaves around criticality while driven dynamically, we introduce following ansatz,
\begin{equation}
\eta=f_4\left(\alpha R^{-1/r\nu}\right), \label{eq:kz_an2}
\end{equation}
where $f_4[.]$ is arbitrary function. Then, by rescaling $\alpha$ as $\alpha R^{-1/r\nu}$, the dynamical curves collapse onto a single universal curve, as shown in Fig.~\ref{fig:kz_defect}(b). Thus, this dynamical scaling collapse provides independent confirmation of the critical exponents extracted from the static analysis.

\section{\label{sec:conc}Conclusion}
We investigate localization phenomena in an interacting Bose gas confined within a tilted OL using complementary nonequilibrium protocols. The model realizes a Stark-induced route to localization distinct from conventional disorder-driven Anderson localization or quasiperiodic Aubry–Andr{\'e} transitions. Sudden interaction and tilt quenches reveal distinct dynamical signatures of localized and delocalized phases through survival probability and spectral analysis. Localized states show persistent revivals and periodic dynamics, whereas delocalized states exhibit broadened spectra and irregular temporal behavior. Taken together, these results provide a unified dynamical characterization of the localization phenomena.

We further examine nonequilibrium ramp dynamics across the transition within the framework of the KZ mechanism. Static finite-size scaling of localization length and Bogoliubov excitation gaps yields critical exponents associated with the localization transition. Using these exponents, driven dynamics across criticality exhibits KZS collapse, confirming universal breakdown of adiabaticity.
  
From an experimental perspective, the KZ scaling discussed here can be tested by preparing the condensate in the localized regime and ramping the tilt strength at different rates. The localization length may be extracted from in situ density profiles through the RMS width. The breakdown of adiabaticity can be inferred from the rate-dependent broadening of the condensate profile or deviations from the adiabatically expected density distribution. A collapse of these quantities with the rescaled ramp-rate would provide a direct dynamical means for verifying the KZ mechanism.

These findings of this work suggest that nonequilibrium dynamics can serve as a sensitive probe of Stark-type localization in interacting Bose gases. Future work may explore the stability of the observed scaling against  finite temperature, atom losses, and external noise, as well as its extensions beyond the applicability regime of the GP treatment.


\bibliography{apssamp}

\end{document}